\documentclass[sigconf]{acmart}
\AtBeginDocument{%
  }

\setcopyright{acmlicensed}

\copyrightyear{2026}
\acmYear{2026}
\setcopyright{cc}
\setcctype{by-nc-nd}
\acmConference[UIST '26]{The 39th Annual ACM Symposium on User Interface Software and Technology}{November 02--05, 2026}{Detroit, MI, USA}
\acmBooktitle{The 39th Annual ACM Symposium on User Interface Software and Technology (UIST '26), November 02--05, 2026, Detroit, MI, USA}
\acmDOI{10.1145/3830398.3830534}
\acmISBN{979-8-4007-2856-3/2026/11}

\usepackage{listings}
\usepackage{enumitem}
\usepackage{framed}
\usepackage{makecell} 
\definecolor{darkgreen}{RGB}{30,150,50}
\definecolor{codegray}{RGB}{240,240,240}

\lstdefinestyle{codeblockstyle}{
  basicstyle=\ttfamily\small,
  breaklines=true,
  breakatwhitespace=false,
  columns=fullflexible,
  keepspaces=true,
  escapechar=|,
  commentstyle=\color{gray},
  backgroundcolor=\color{codegray},
  frame=single,
  framerule=0.8pt,
  rulecolor=\color{black},
  framesep=6pt,
  xleftmargin=6pt,
  xrightmargin=6pt
}
\lstnewenvironment{codeblock}{\lstset{style=codeblockstyle}}{}

\usepackage{algorithm}
\usepackage{algorithmic}
\usepackage{soul}
\newif\ifcleanversion
\cleanversiontrue   
\newcommand{\changed}[1]{%
  \ifcleanversion
    #1%
  \else
    \textcolor{blue}{#1}%
  \fi
}

\begin{document}

\title{ShadAR: Real-time Prompting On-demand Visual Augmentation in Augmented Reality}

\author{Yanni Mei}
\email{yanni.mei@tu-darmstadt.de}
\affiliation{%
  \institution{TU Darmstadt}
  \city{Darmstadt}
  \country{Germany}
}

\author{Samuel Hartenstein}
\email{samuel.hartenstein@stud.tu-darmstadt.de}
\affiliation{%
  \institution{TU Darmstadt}
  \city{Darmstadt}
  \country{Germany}
}

\author{Jonas Wombacher}
\email{jonas.wombacher@tu-darmstadt.de}
\affiliation{%
  \institution{TU Darmstadt}
  \city{Darmstadt}
  \country{Germany}
}

\author{Florian Mueller}
\email{florian.mueller@tu-darmstadt.de}
\affiliation{%
  \institution{TU Darmstadt}
  \city{Darmstadt}
  \country{Germany}
}

\author{Jan Gugenheimer}
\email{jan.gugenheimer@tu-darmstadt.de}
\affiliation{%
  \institution{TU Darmstadt}
  \city{Darmstadt}
  \country{Germany}
}

\renewcommand{\shortauthors}{Mei et al.}

\begin{abstract}

Augmented Reality can reshape our visual perception (e.g., simulating color blindness) yet such effects are typically pre-programmed by experts, making visual augmentation static and inflexible.
We introduce ShadAR, an AR system that allows users to describe expected visual augmentations (e.g., "Blur all people") for changing their vision in real time. 
ShadAR can graphically manipulate AR passthrough real-time via shader injection. 
When receiving a user’s prompt, ShadAR uses LLMs (GPT5.2) to interpret the user’s intent, analyzes the user’s view with object detection (YOLOv26), generates a pixel shader, and applies it to the targeted region on AR passthrough.
To understand usage pattern, we invited 21 HCI researchers to use ShadAR, identifying four application domains (design, health, productivity, and social interaction) and two prompt-framing patterns: \textbf{procedural prompts} specifying what to modify and how (\textit{``Highlight the book.''}), and \textbf{end-goal prompts} conveying desired mental/behavioral states (\textit{``Make me laugh.''}). 
Such end-goal prompts suggest a future in which users specify outcomes (e.g., \textit{``Make me not miss my flight''}) while the system determines how to realize them, though often with inherently indeterminate results. 
We discuss design opportunities for ShadAR and ethical concerns.





\end{abstract}


\begin{teaserfigure}
  \centering
  \includegraphics[width=1\textwidth]{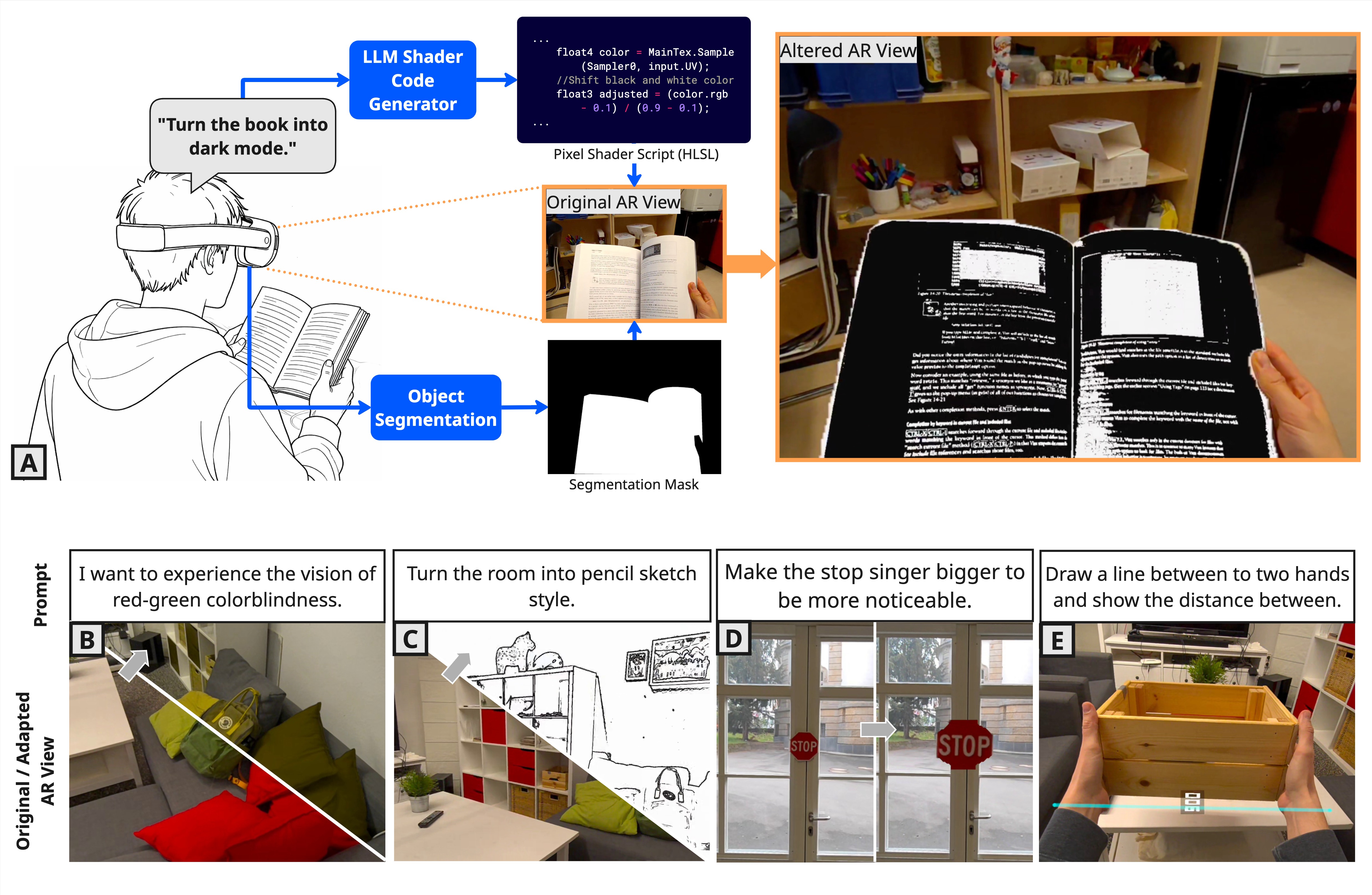}
  \caption{ShadAR lets AR users verbally request visual augmentations (e.g., "Turn the book into dark mode") and uses LLMs to generate shaders with object detection to modify their view. Other examples: (B) simulating color blindness; (C) applying a sketch-style filter to the room; (D) enlarging a stop sign; and (E) drawing a line between two hands to show the distance}
  \Description{}
  \label{fig:teaser}
\end{teaserfigure}


\maketitle

\section{Introduction}


Visual perception is a central interface between a person and the real world. 
Augmented Reality (AR) can intervene in this interface by computationally modifying the passthrough feed that mediates how users see the world, known as ``visual augmentation'' \cite{langlotz2024design}.
Prior work explored different use cases of AR visual augmentation, such as recoloring real-world objects to compensate for colorblindness \cite{flatla2015colourid}, or blurring distractions to support focus \cite{hong2024visual}.
However, these visual effects are predefined by experts for specific usages. 
While effective in their intended scenarios, this approach might not suit the diverse and unpredictable needs that arise in future \textit{EverydayAR}\cite{tran2025wearable}, where people use AR in many different contexts and tasks. 
During a day, a user may require various visual augmentations, ranging from increased contrast in low-light conditions, to edge enhancement during manual work, to blurred distractions for reading. 
To address such diverse needs, it requires visual augmentations to be created and adjusted by users on demand and in real time. 
However, users lack the technical skills to create such augmentations, as this typically involves writing computer graphics code, such as shader.



Recent advances in LLM-based code generation suggest a promising solution. 
LLMs can create executable code from natural language descriptions. Prior work has applied this capability in XR to generate C\# scripts that manipulate virtual objects, create application logic, or adapt narrative contents based on user input \cite{de2024llmr, giunchi2024dreamcodevr, zhu2025agentar}.
In contrast, rather than using code generation to modify virtual content external to the user, our work takes an ``egocentric approach''. 
We use LLMs to generate HLSL shader code that directly changes how users perceive the real world.


In this paper, we introduce ShadAR, an AR system that enables users to verbally describe the desired visual augmentations (e.g., \textit{``Highlight my book''}) and modify their view at runtime.
ShadAR runs on the Quest 3 with a connected desktop server and can render and graphically manipulate the AR passthrough feed via shader injection. 
When receiving user prompts, the system transcribes audio via \textit{Whisper}, uses an LLM (\textit{GPT-5.2}) to interpret user intent and generate pixel shader code, and segments relevant objects in the user’s view using \textit{YOLOv26}.
It then compiles the shader code at runtime and applies it to the targeted region of the AR passthrough feed, modifying the user’s view as requested.
Therefore, the system enables users to create visual augmentations without writing shader code themselves.


First, we conducted a technical evaluation of ShadAR using 30 prompts under low and high reasoning settings of GPT-5.2.
The overall duration from speech completion to shader deployment was $32.67 \pm 10.13$ seconds with low reasoning and $42.66 \pm 13.80$ seconds with high reasoning. Once generated, the shaders ran at 44.1 FPS and 45.5 FPS, respectively.
Of the 30 generated shaders, 26 were syntactically correct after a single prompt with low reasoning, and 28 with high reasoning.

Next, to understand the application and usage patterns of ShadAR, we conducted an ideation workshop with HCI researchers (N = 21) to identify potential application scenarios and to analyze how users phrase requests for on-demand visual augmentation.
Using thematic analysis of their prompts and shaders, 
we discovered two prompt-framing patterns: \textbf{procedural prompts}, which specify targeted objects and visual transformations (e.g., \textit{``Pixelate everything except my screen.''}), and \textbf{end-goal prompts}, which convey a desired visual, mental, or behavioral state (e.g., \textit{``Make me laugh.''}).
These patterns differ in output determinacy: procedural prompts constrain LLM interpretation and produce more deterministic results, whereas end-goal prompts leave greater room for LLM interpretation, producing more diverse and unpredictable outcomes.
This suggests implications for the future design of ShadAR: prompting visual augmentation should support not only direct generation, but also interaction for user clarification, offering alternative results, and enabling iterative co-exploration when user intent is unclear.


We argue that such end-goal prompts could become a key usage pattern for end users to request visual augmentations for everyday uses (e.g., \textit{``Remind me to water my plant.''}), as they reduce the effort and skills needed to create step-by-step visual instructions. 
At the same time, they leave a large interpretation space for LLMs. This opens up design opportunities such as integrating personal preferences and contextual information for better intent interpretation, and adding explainability by indicating how LLMs select a visual solution to fulfill the user's goals.
\changed{The whole source code is available at GitHub \footnote{\url{https://github.com/teamdarmstadt/ShadAR}}
}


In this work, we present two main contributions:
\begin{itemize}
    \item The technical implementation of ShadAR, including shader code generation, real-world segmentation, run-time shader compiler, and selective shader rendering.
    \item The empirical insights from a workshop with 21 HCI researchers regarding new application scenarios and usage patterns (procedural prompts, end-goal prompts)
\end{itemize}

\section{Related Work}

In the related work section, we review visual augmentation in AR, the use of LLMs for generating computer graphics scripts, and the current state of LLM-based code generation in XR.

\subsection{Visual Augmentation in AR}

AR is not limited to superimposing virtual content to the real world; it can also change how the real world itself is visually perceived by users, known as ``visual augmentation'' \cite{langlotz2024design,sutton2019computational}.
With recent advances in video passthrough AR, the user’s view is reconstructed from camera input, enabling more flexible visual augmentation than optical see-through displays \cite{langlotz2024design}.
Computer graphics algorithms, such as shaders, can modify the visual presentations of the real world displayed in the AR passthrough, by changing visual properties such as color, contrast and spatial distortion \cite{langlotz2024design,hong2024visual}.
Visual augmentation has been explored for various uses, including simulating visual deficiencies (e.g., cataract) for accessible design \cite{krosl2020cataract,radler2025vip,abraham2024simulation}, supporting color identification for color-blind users by recoloring objects that are hard to distinguish \cite{langlotz2024design,flatla2015colourid}, and dimming irrelevant information to improve concentration \cite{hong2024visual,katins2025ad}.

However, visual perception needs vary widely depending on context and personal preference. 
For example, some users prefer edge highlighting to guide attention, but in visually dense environments, this can become fatiguing, leading them to choose contrast enhancement instead \cite{moshtael2015high,satgunam2012effects}. In privacy-preserving visualization, users' trust also differs across techniques such as blur, pixelation, and semantic removal \cite{offermann2023shapes,cruz2025assessing}. 
These differences highlight the need for users to create visual augmentations based on their immediate needs and preferences. 
However, most users lack the expertise to write shader code. 
Recent advances in LLM-based code generation offer opportunities to take over shader writing and support users in creating visual augmentations.

\subsection{LLMs Computer Graphic Code Generation}

Using LLMs to generate computer graphics (CG) scripts (e.g., Processing\footnote{\url{https://processing.org/}}, OpenGL\footnote{\url{https://learnopengl.com/Getting-started/Shaders}}) has been widely explored in creative coding, supporting artists to produce expressive graphics \cite{angert2023spellburst,wang2026reflexa,yuksel2025ai,moon2024teaching,wu2024exploring}.
CG scripts, such as shaders, directly manipulate pixel properties (e.g., RGB values, spatial transformations) to create visual effects, motion and abstract shapes \cite{mcreynolds2005advanced}. 
Compared to AI image generation \cite{liu2022design}, CG code is less versatile but more explicitly parameterized, allowing precise control and efficient real-time rendering \cite{mccool2002shader,pharr2008programmable}.
Prior work mainly focuses on translating artists' natural language prompts into CG code. 
One challenge is mapping high-level visual concepts to low-level implementations \cite{yuksel2025ai}.
For example, Spellburst shows the difficulty of translating descriptions like “wavy ocean” into code (e.g., sine functions, color choices), and addresses this using few-shot prompting and chain-of-thought \cite{angert2023spellburst}. 
Although rapid advances in LLMs' reasoning abilities have alleviated many of these challenges, domain-specific applications still require additional effort. 
In our case, translating desired AR visual augmentation into shader code requires instructions for the LLM to understand user intent, scaffolding shader generation for AR passthrough modification, and integrating external data like object detection.

Besides application-specific technical challenges, most prior work in LLM-based CG code generation focuses on desktop tools for artistic expression \cite{angert2023spellburst,wang2026reflexa}. 
In contrast, we explore a practical use case of modifying visual perception in everyday scenarios, and we focus on a new medium --- XR. 
Next, we introduce how XR is combined with LLM-based code generation in current practices.


\subsection{LLMs Code Generation in XR}

LLMs' code generation is being explored in XR to generate system-operation code (C\#) for tasks like manipulating 3D content \cite{de2024llmr,giunchi2024dreamcodevr,kurai2025magicitem}, building interactive applications \cite{zhu2025agentar,wang2024virtuwander}, or adapting narratives \cite{wu2024artist,dogan2024augmented}.
For example, LLMR generates C\# code to manipulate objects reacting to users' prompts (e.g., \textit{``Make the bear twice as large.''}), or create new scenes \cite{de2024llmr}. 
Such tasks are limited to VR, because only virtual content can be changed.
In AR, code generation can build applications by creating application logic, interfaces, and narrative content.
For instance, AgentAR generates AR applications which can offer situated guiding information to support brewing coffee \cite{zhu2025agentar}.
Across both VR and AR, prior work focuses on \textit{exocentric modifications}, by generating system-operation code (e.g., C\#) to change virtual content external to users.
In contrast, ShadAR focuses on \textit{egocentric modifications}, directly transforming the user’s own vision by generating shaders that operate at the final image stage.

\section{ShadAR - User Interaction}
\begin{figure*}
    \centering
    \includegraphics[width=1\linewidth]{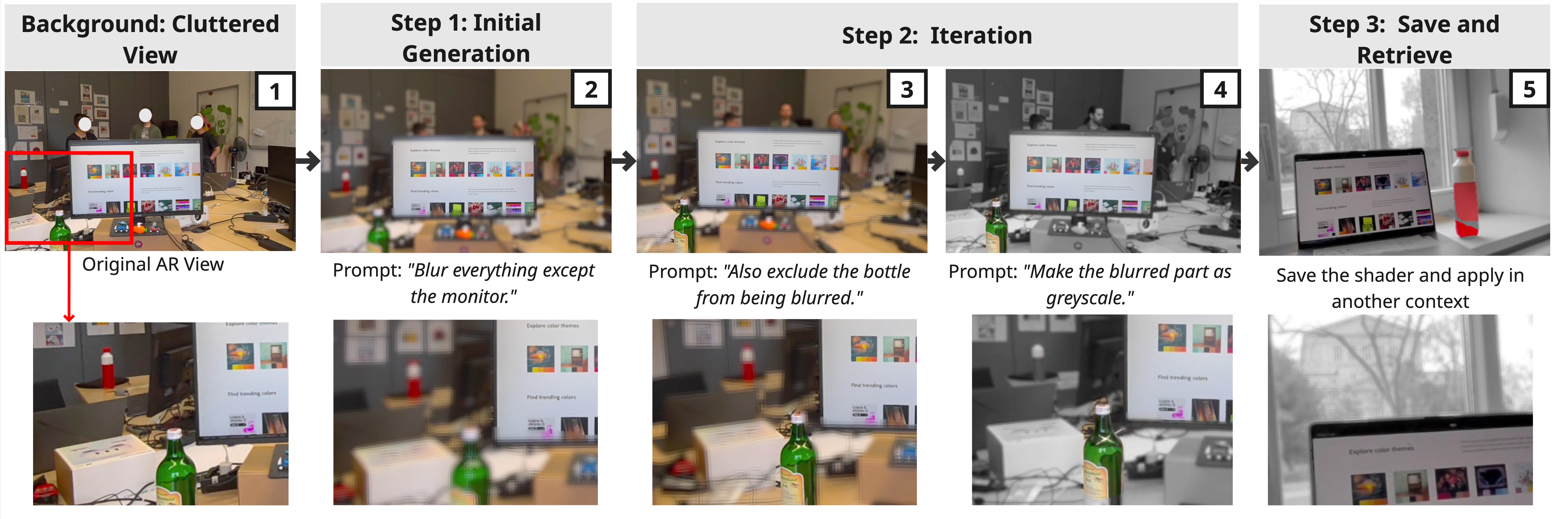}
    \caption{The scenario of Anne using ShadAR to de-emphasize clutter in the office to enhance concentration. To note, in (1) white circles were added by authors for anonymity, and were not generated by ShadAR. }
    \label{fig:scenario}
\end{figure*}
We present ShadAR, an AR system that lets users iteratively author visual augmentations of the passthrough view through natural-language prompts.
To use ShadAR, a user wears a Quest 3 headset and directly speaks their desired visual effect in natural language (e.g., \textit{``Blur everything except my screen''}). 
The AR view will be adapted according to the user’s requests.
Users can also iterate existing visual augmentations by continuing to speak commands (e.g., \textit{``Increase blurriness''}).
The visual augmentations can be saved and retrieved for future use.

Below, we illustrate how a user, Anne, uses ShadAR to adjust her vision for concentrating on work (Fig. \ref{fig:scenario}).

\textbf{Background} One day, Anne was working at her monitor in a messy office with people moving around, making it difficult to concentrate (Fig.\ref{fig:scenario}-1).
She decided to use ShadAR to solve the problem. 

\textbf{Step 1: Speaking Out Prompt to Generate Visual Augmentation} 
First, she pressed the controller button to record audio and spoke: \textit{``Help me focus on my monitor, blur other things.''} 
After that, she saw the AR view changed as she requested (Fig.\ref{fig:scenario}-2).

\textbf{Step 2: Iterate Visual Augmentations}
She found that the bottle should not be blurred, because otherwise she might forget to drink. So she said: \textit{``Also exclude the bottle from being blurred.''} (Fig.\ref{fig:scenario}-3)
Then, she still wanted the clutter to be less distracting, so she spoke \textit{``Make the blurred part as grayscale.''} (Fig.\ref{fig:scenario}-4) 

\textbf{Step 3: Save and Retrieve Shaders}
She was happy with the result and stored it on the headset as a focus mode for work for future use (Fig.\ref{fig:scenario}-5).

\section{Technical Architecture}

\begin{figure*}
    \centering
    \includegraphics[width=\linewidth]{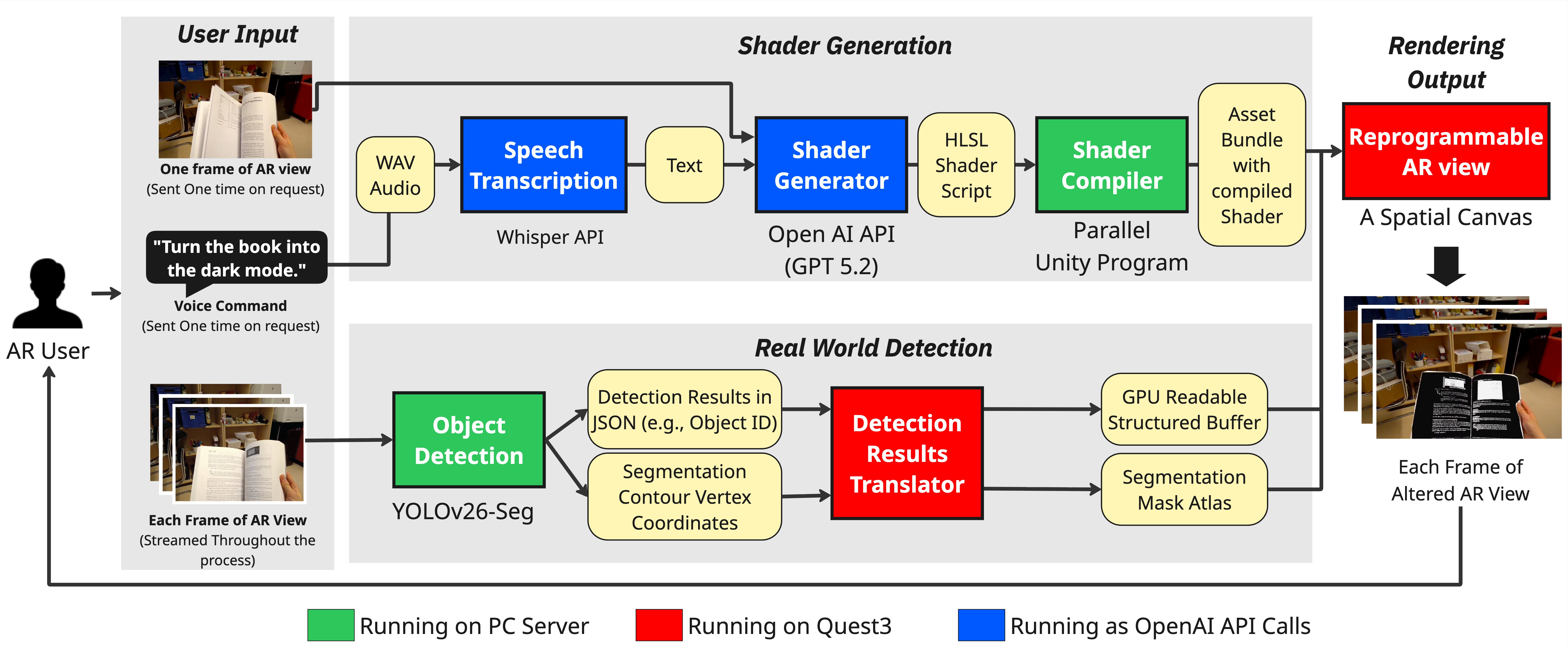}
    \caption{The technical pipeline of ShadAR.}
    \label{fig:tech}
\end{figure*}

ShadAR addresses three technical challenges: translating natural-language requests into executable pixel shaders, grounding those shaders in the current physical scene through object detection and segmentation, and deploying them at runtime despite Quest passthrough and Unity compilation constraints. To solve these challenges, we partition the system across headset, desktop server (RTX 5090), and external APIs, yielding a six-stage pipeline (see Fig. \ref{fig:tech}). Below, we illustrate the workflow of the pipeline.

First, when the system receives a user’s voice prompt (e.g., \textit{``Turn the book into dark mode''}), it uses \textcolor{blue}{\textbf{Speech Transcription}} (via a Whisper API call) to convert the audio into text. 
This transcript, along with a frame of the user’s current AR view, is sent to the \textcolor{blue}{\textbf{Shader Generator}} (via an OpenAI API call with a specifically designed system prompt allowing the LLM to reference detected objects), which interprets the user’s intent and generates a pixel shader. The generated shader script is sent to the \textcolor{darkgreen}{\textbf{Shader Compiler}} (a running-in-parallel Unity Instance) on the desktop, which compiles it at runtime and then sends it to the Quest.
To handle prompts referencing real-world objects (e.g., a book), ShadAR streams each AR passthrough frame to the \textcolor{darkgreen}{\textbf{Object Detection}}, which is running YOLOv26 on the desktop, to recognize and segment objects in the user's AR view.
The results are streamed back to the Quest 3, where the \textcolor{red}{\textbf{Detection Result Translator}} converts them into structured buffers and mask atlases that the shader can read and use. 
Finally, the compiled shader and detection results are passed to the \textcolor{red}{\textbf{Reprogrammable AR Renderer}}, a spatial canvas that renders AR passthrough outside the Quest’s default pipeline. The renderer applies the shader in real time to targeted regions (e.g., a book) in the AR passthrough texture, so as to modify the user’s view.

Next, we will elaborate on the key technical components.

\subsection{Real World Detection}
\changed{
Real-world detection runs a YOLO26 segmentation
model\footnote{\url{https://docs.ultralytics.com/models/yolo26/}} on the
desktop throughout the AR experience, identifying and segmenting objects in
the user's AR view in real time. Each frame of the AR passthrough video is
streamed from the Quest 3 to the desktop for analysis, and the detection
results are sent back to the headset, comprising (1) object recognition data
in JSON format (e.g., object ID, confidence score) and (2) segmentation
binary masks.
\textbf{\textit{Challenge: }}
Because detection results are streamed every frame, the payload must be kept
small for efficient transmission. Segmentation masks, however, are images and
are correspondingly large.
\textbf{\textit{Solution: }}
Instead of streaming the full mask image, we convert the contour of each
segmented object into a 64-vertex polygon and transmit only the normalized 2D
coordinates of its vertices. This reduces data size while preserving shape
fidelity.
}

\subsection{Detection Results Translator}
When the headset receives recognition and segmentation results, the system must convert this data into a format which the shaders can read and use. 

\subsubsection{Convert Object Recognition Data to Struct Buffer}
The object recognition data (e.g., LabelID, Confidence Score, Bounding Box Coordinates) is sent to the Quest in JSON format, but shaders cannot directly read JSON. 
Therefore, we parse the data into a custom C\# \textit{struct} called \textit{Detection} for each detected object, and store it in a \textit{Structured Buffer} called \textit{Detections} that shaders can read.

\subsubsection{Convert Segmentation Polygon Coordinates to Mask Atlas}


On the headset, segmentation contour coordinates are rasterized back into a 2D binary mask texture. 
To improve rendering efficiency when handling multiple objects in one frame, all individual masks of each detected object are packed into a single mask atlas -- a 2D texture organized as a 4 × 4 grid, where each cell stores one mask.
Shaders can access this atlas to retrieve relevant masks.


\subsection{Shader Generator}
We use ChatGPT 5.2 with a designed system prompt as the Shader Generator.
It receives the transcribed prompt and one frame of the user's AR view, interprets the user's intent, and then generates the shader script.

\subsubsection{Shader Generation Scaffolding}

\textbf{\textit{Goal: }}We need to ensure the LLM can (1) translate the user's intent into low-level shader code and (2) reduce syntax errors. Also, the resulting shader can (3) directly modify the AR passthrough and (4) can read and use YOLO results for selective rendering. 
\textbf{\textit{Solution: }}We achieve these 4 requirements through the design of a system prompt (Appendix.\ref{sec:appendix_shader}):

\changed{
(1) Users provide high-level requests about a desired look
(e.g., \textit{``Turn the book into dark mode.''}) rather than low-level
shader programming instructions, so we need to ensure the LLM can
\textbf{translate user intent into shader generation tasks}. To support
this, we send the user's prompt together with a screenshot of their current
AR view. The system prompt instructs the LLM that its task is to modify the
visual perception represented in that screenshot by generating a shader
that overlays graphical effects on it. In our experiments, we found that
GPT-5.2 can infer basic visual strategies from prompts without explicit
examples (e.g., translating "dark mode" into a shift between white and
black).
}

(2) To \textbf{improve syntactic correctness} for shader generation, we provide a list of shader structure specifications and common mistakes to avoid in the system prompt. This list was derived by first generating shaders using GPT-5.2 and analyzing recurring syntax errors.

(3) To ensure \textbf{shaders apply visual effect on top of the AR passthrough feed}, we scaffold the shader script in three key steps: (a) use the AR video stream as the base texture, (b) only modify pixel parameters of this base texture (c) only return the modified base texture. 
These three steps ensure that the generated shader transforms the AR passthrough by modifying original pixels, rather than replacing the AR passthrough by covering it with entirely synthetic pixels.
These three steps are elaborated in a code snippet with explanatory comments, which is included in the system prompt for one-shot learning.

\changed{
(4) To ensure \textbf{the generated shader knows how to read and use the
detection results}, and can selectively apply effects to the objects named
in the user's prompt, we scaffold shader generation with three mechanisms.
\textbf{(a) Label ID Mapping.} We include the YOLO object
list~\cite{lin2014microsoft} in the system prompt and instruct the LLM to
map the user's prompt to the corresponding YOLO label ID (e.g., "highlight
the crowd" $\rightarrow$ people $\rightarrow$ labelID~=~0), which is written
directly into the generated shader for later filtering.
\textbf{(b) Custom Data Struct Declaration.} Detection results are passed as
a custom struct. We include its definition in the system prompt and instruct
the LLM to declare the \texttt{Detection} struct in any shader that involves
object detection, so the results can be read.
\textbf{(c) Filtering Logic.} We enforce a filtering rule so that effects
apply only to pixels belonging to the target object. For each pixel $p$, we
evaluate all Detection structs $d$ and select $p$ if (i) the detection
carries the target label, (ii) $p$ lies within the bounding box $B_d$, and
(iii) the segmentation mask $M_d$ marks $p$ as part of the object. This
logic (Algorithm~\ref{alg:pixel_mask}) is provided in the system prompt as a
code snippet for one-shot learning.
}

\subsubsection{Memory management for iterating or redoing shader}
\changed{
ShadAR supports both iterating on an existing shader and starting from
scratch, by controlling the chat history in the OpenAI API call. For
iteration, the previous prompt and generated shader remain in the
conversation context, enabling incremental refinement through new prompts
(e.g., adjusting parameters only). For a complete redo, users clear the
chat history before submitting a new prompt.
}

\begin{algorithm}[H]
\caption{Filtering Logic For Selective Rendering}
\label{alg:pixel_mask}
\begin{algorithmic}[1]
\FOR{each pixel $p = (x,y)$}
    \FOR{each Detection $d$}
        \IF{$d.\text{labelID} = \text{targetID} \land p \in B_d \land M_d(p) = 1$}
            \STATE applyEffect($p$)
        \ENDIF
    \ENDFOR
\ENDFOR
\end{algorithmic}
\end{algorithm}

\subsection{Real-time Shader Compiler}
When receiving the generated shader script, the system needs to compile it so that the shader can be rendered during runtime.
\textbf{\textit{Challenge: }}
However, in Unity applications, shader code cannot be compiled at runtime.
\textbf{\textit{Solution: }}
To address this limitation, we run a second Unity instance in parallel on a desktop to receive and compile the shader code. 
To accelerate compilation, the desktop Unity instance runs in headless mode\footnote{\url{https://docs.unity3d.com/6000.3/Documentation/Manual/desktop-headless-mode.html}}, executing compilation functions through the command line without initializing the graphical user interface.
Then, the compiled shader is packaged into an \textit{asset bundle}\footnote{\url{https://docs.unity3d.com/6000.3/Documentation/Manual/AssetBundlesIntro.html}} and transmitted to the Quest. This asset bundle can then be loaded at runtime, allowing us to inject shader code into a running Unity application.


\subsection{Reprogrammable AR View}

\changed{
The compiled shader and object detection results feed into the
Reprogrammable AR View, which graphically manipulates AR passthrough in
real time via shader injection.
\textbf{\textit{Challenge: }}
Quest 3 does not permit access to or modification of its original
stereoscopic rendering.
\textbf{\textit{Solution: }}
We therefore render the video passthrough independently, replacing the
Quest's own. We first deactivate the Quest's default passthrough rendering,
then capture the passthrough feed (1280\,$\times$\,960) from the built-in
camera via the \textit{Meta Passthrough Camera
API}\footnote{\url{https://developers.meta.com/horizon/documentation/spatial-sdk/spatial-sdk-pca-overview/}}.
Because Meta exposes only one of the two external cameras at runtime, we
create a monoscopic AR view by rendering the passthrough onto a 2D canvas
(1280\,$\times$\,960 pixels). The feed is imported into Unity as a
\textit{WebCamTexture} and embedded in the shader of the canvas material.
\textbf{By dynamically replacing the shader assigned to this material, the
AR passthrough rendering becomes graphically manipulable.}
To approximate a plausible monoscopic AR perspective, the canvas holds a
fixed spatial relationship to the \textit{LeftEyeAnchor} in Unity: its X and
Y coordinates align with the center of the eyes, and it remains
perpendicular to the eye's forward direction at a fixed Z offset.
}

\section{Technical Performance Evaluation}
We evaluated the technical performance of the ShadAR pipeline. 
Experiments were conducted on a desktop server (RTX 5090, Dedicated VRAM: 31.5 GB, Shared VRAM: 72.6 GB), connected via Wi-Fi (1 Gbps), and deployed to a Meta Quest 3 headset. 
To benchmark the system, we generated 30 prompts using Claude, ensuring that all prompts were within the capabilities of the YOLO object detection.
We tested in 2 conditions of both low and high reasoning settings of GPT5.2, and each prompt was tested under both settings.

We measured end-to-end latency, defined as the time from speech completion to the application of visual effects in the AR view. This includes audio encoding, speech transcription, shader generation, shader compilation, bundle transmission, and deployment to the AR device.
In addition, we evaluated rendering performance by measuring per-frame processing time during the 5-second interval following the application of shaders. Notably, we did not measure the overall Unity rendering performance, but specifically the video frame update time on the spatial canvas.

We report syntax error rate, end-to-end latency, code generation latency for successfully compiled shaders, and FPS in Table~\ref{tab:technicalEvaluation}. In the low and high reasoning settings, approximately 72.6\% and 81.1\% of the total latency is spent on OpenAI-based code generation, respectively.
We plan to reduce this overhead in future work by using a \changed{locally deployed LLM} for shader generation.

\changed{
Additionally, we measured the latency of the object detection, which is the round-trip time (RTT) from when the headset transmits a frame to when it
receives the corresponding detection result, under four conditions containing one, two, three, and four objects in the view frame, each recorded over a one-minute interval. 
Mean RTT is: $23.08 \pm 4.13$, $27.39 \pm 5.56$, $30.17 \pm 8.29$, and
$31.18 \pm 5.94$~ms for one through four objects, respectively, suggesting that more complex scenes incur higher detection latency. Because a new frame is sent only after the previous detection result is received, this latency introduces a temporal offset between the segmentation mask and the passthrough video.
}

\changed{
The motion-to-photon latency includes two parts: video capture latency from the Passthrough Camera API, which Meta reports as 20–40 ms \cite{metaPCA}, and render processing time on the 2D canvas, which we did not specifically measure.
}


\begin{table}[t]
\centering
\small
\begin{tabular}{lcccc}
\hline
\makecell{\textbf{Reasoning}\\\textbf{Setting}}
& \makecell{\textbf{Syntax}\\\textbf{Error}} 
& \makecell{\textbf{Overall}\\\textbf{Latency (s)}} 
& \makecell{\textbf{Code Gen.}\\\textbf{Latency (s)}} 
& \makecell{\textbf{FPS}} \\
\hline
Low  & 4 & $32.67 \pm 10.13$ & $23.73 \pm 8.39$ & $44.1$ \\
High & 2 & $42.66 \pm 13.80$ & $34.60 \pm 13.74$ & $45.5$ \\
\hline
\end{tabular}
\caption{Technical performance of ShadAR.} 
\label{tab:technicalEvaluation}
\end{table}




    

    
    
    

\section{Ideation Workshop with HCI Researchers}
We conducted an ideation workshop with HCI researchers to explore the design space of user-authored visual augmentations and to understand how users formulate requests for such augmentations. We deliberately recruited HCI researchers with experience in research prototypes to enable a broader and more reflective exploration of ShadAR’s design space. Accordingly, we planned the workshop not to estimate real-world usability, but to surface potential application domains and prompt formulations that this new class of system enables.

\subsection{Participants}
\changed{
We invited 21 HCI researchers (M=10, F=7, Other=1, Prefer Not To Say=3, with a mean age of 30.67±7.53) in domains of interaction design, AR/VR and Human-AI interaction.
All participants had experience with research prototypes.
Participants were divided into 5 groups (G1-G5).
}

\subsection{Procedure}
\changed{
The session lasted approximately 90 minutes with the whole 5 groups in parallel, with each group (4–5 people) sharing one headset and testing prompts with the live Quest prototype. The session followed three stages:
}

\textbf{Step 1: Introduction and Demonstration} 
The first author introduced the ShadAR concept by presenting demo videos and explained the procedure and tasks in the workshop.

\textbf{Step 2: Initial Prompt Ideation} 
We asked participants to imagine a future in which everyday AR is available to them, and ShadAR could be used in everyday contexts. We invited participants to brainstorm prompts for changing visions and their purpose, and write them down in a Miro board.
We encouraged participants not to restrict themselves to the current technical ability of ShadAR, regarding the shaders' visual transformation ability and YOLO detection scope.

\textbf{Step 3: Ideation with Prototype Exploration} 
After the initial ideation, participants selected prompts they thought were feasible with the current prototype, tested them using ShadAR, and added new ideas.
They were informed about the graphic manipulation ability of shaders and the detection abilities of the YOLO model.

\subsection{Data Analysis}
\changed{
We collected two sets of data: (1) the prompts and participants' explanations of their usages, and (2) the generated shaders. 
In total, the participants generated 63 prompts, of which 26 were used to create shaders with ShadAR. 
We conducted a thematic analysis of the generated prompts, together with watching the visual effects of their shaders.
Two researchers coded the data and iteratively grouped codes into themes. 
}

\subsection{Application Scenarios}
\label{subsec:scenarios}

\begin{figure*}
    \centering
    \includegraphics[width=1\linewidth]{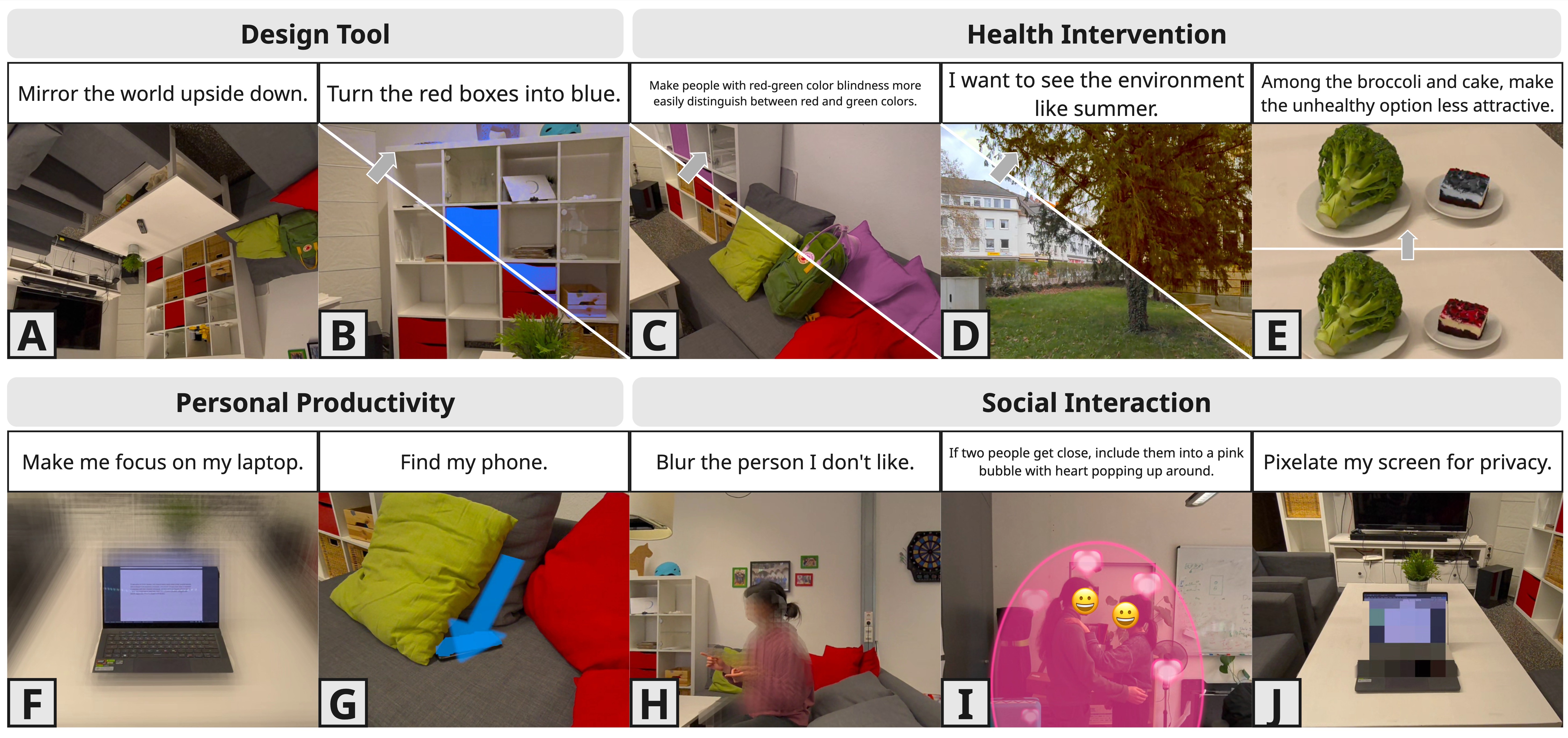}
    \caption{Scenarios of ShadAR generated from the workshop, including design, health, productivity, and social interaction. To note, in (I), the emoji icons were added by authors for anonymity, and not generated by ShadAR.}
    \label{fig:usecases}
\end{figure*}

Through the prompts and participants' explanations of their usages, we discovered 4 application scenarios of ShadAR.

\textbf{Scenario 1: Design Toolkit.}
We discovered 20 prompts using ShadAR as a design tool to support inclusiveness (8), creativity (8) and quick idea realization (4).
The first usage is as an \textbf{Empathy Tool} for designers to simulate visual impairments (e.g., colorblindness, tunnel vision) and support accessible design. For example, G2 proposed \textit{``See the world through the eyes of a dog.''}, to experience the visual ability of dogs, which are red-green colorblind.
The second usage is as a \textbf{Creative Lens}, offering a novel or artistic filter to inspire creativity. G1 mentioned \textit{``I want to see the environment like a Ghibli studio movie.''}. G4 prompted an unconventional vision: \textit{``Mirror the world left and right and up and down''} (Fig.\ref{fig:usecases}-A).
The last usage is for \textbf{Rapid Design Idea Externalization}, such as recoloring (Fig.\ref{fig:usecases}-B) or adding graphical patterns on products, as G2 put it \textit{``Make the wall turquoise and the sofa in the complementary color.''}

\textbf{Scenario 2: Health Intervention.}
We discovered 12 prompts using ShadAR as a health intervention tool for visual correction (3), mental wellbeing (6) and healthy behavior nudging (3).
First, ShadAR can be used as a \textbf{Visual Deficiency Compensation} tool. One participant from G2, who has red-green colorblindness, prompted \textit{``Make people with red-green color blindness more easily distinguish between red and green colors.''} and ShadAR recolored the red objects toward pink to distinguish from green (Fig.\ref{fig:usecases}-C). This actually enabled him to discover a green hoodie of another participant, which he had not discovered before.
Second, some prompts aim to modulate emotion for supporting \textbf{Mental Health}. G1 prompted \textit{``I want to see the environment like summer.''} (Fig.\ref{fig:usecases}-D) for dealing with winter depression. 
Another example is \textit{``Make me laugh.''} from G1. This triggered ShadAR to create funny face distortion effects.
Finally, participants also used ShadAR to \textbf{Nudge Healthy Behavior} (Fig.\ref{fig:usecases}-E). G5 mentioned \textit{``Highlight all vegetarian products and blur all other products.''} for guiding grocery choices.
G2 prompted \textit{``Make all displays fully black over time.''} to limit screen time.

\textbf{Scenario 3: Personal Productivity.}
There were 15 prompts using ShadAR to optimize productivity in daily life, such as supporting concentration (7), visual searching (6), or offering superhuman vision (2).
First, ShadAR can be used to remove visual noise \cite{hong2024visual} for \textbf{Enhancing Concentration}.
For instance, G3 prompted \textit{``Focus mode on the screen.''} (Fig.\ref{fig:usecases}-F).
Second, some prompts aimed at speeding up \textbf{Visual Searching}. For example, G4 created prompts \textit{``Help me find the phone.''} (Fig.\ref{fig:usecases}-F) and \textit{``Make a green frame around the emergency exits.''}.
Finally, people expected \textbf{Superhuman Vision} \cite{langlotz2024design} to extend visual ability to approach complex problems. Examples were \textit{``While playing sports, help me show where the ball will go or where I need to stand.''} (G5) or having \textit{``Night Vision.''} (G3) in a dark alleyway.

\textbf{Scenario 4: Social Interaction.}
We found 12 prompts using ShadAR in social contexts, to fix social pressure (7), support personal expression (3) or enhance privacy (2).
First, some prompts used ShadAR to \textbf{Fix Social Stress}. G5 prompted \textit{``Blur people I don't like.''} to fix unpleasant social confrontation (Fig.\ref{fig:usecases}-H). G1 mentioned augmenting the surrounding people with nude colors for funny effects and to resolve anxiety.
Second, some prompts proposed creative AR lenses to support \textbf{Personal Expression}. G5 asked to overlay weather effects (e.g., snowing) or other expressive graphics to convey one's mood (Fig.\ref{fig:usecases}-I).
Third, some prompts aimed for \textbf{Privacy} in a Social AR context, by regulating people's visual access to sensitive information, such as blacking out \textit{``people's eye''} (G4) and \textit{``screens''} (G3) (Fig.\ref{fig:usecases}-J).

\subsection{Prompt Framing and Output Determinacy}

We identified two closely related properties of prompting: \textbf{Prompt Framing}, referring to whether users described a procedure or an intended end goal, and \textbf{Output Determinacy}, referring to how tightly a prompt constrained the resulting visual effect. These properties were related but not fully independent, as procedural prompts often narrowed the interpretation space, whereas end-goal prompts more often left it open. However, some procedural prompts remained underspecified, while some end-goal prompts implied relatively stable visual operations. In the following, we present our results alongside these axes.

\textbf{Dimension 1: Prompt Framing. }
This dimension captures what users request in their prompts. We discovered two categories.
First, \textbf{Procedural Framing}. This category refers to the user specifying the (1) \textit{\textbf{targeted object}} to be visually altered (e.g., the whole environment, people, object) and (2) \textit{\textbf{visual transformation rules}}, which include recolor (\textit{``Make this sofa in red color''}), change size (\textit{``Make the sign bigger ''}), blur (\textit{``Blur people I don't like.''}), pixelation (\textit{``Pixelate untidy things.''}), distortion, image mirroring (\textit{``Mirror the world upside down''}), vector shape augmentation (\textit{``Add green frame on the emergency exit''}), and style transfer (\textit{``Make the world look like Ghibli films''}). 
Second, \textbf{End-Goal Framing}. 
This category refers to prompts in which users specify the outcome they want to achieve, rather than the visual transformation itself.
The intended outcomes have different aspects including: \textbf{visual} (\textit{``Make me have tunnel vision''}), \textbf{cognitive} (``Improve my focus''), \textbf{emotional} (`\textit{`Make me laugh''}) or \textbf{behavioral} (\textit{``Help me find my phone''}). 
These end-goal prompts do not explain how the LLM can achieve the desired outcome; instead, they rely on the LLM's interpretation to find a possible solution. While these types of prompts currently sometimes result in unwanted outcomes, we believe they reflect the thought process of end users, who are not interested in how to achieve something but in a specific outcome.

\textbf{Dimension 2: Output Determinacy.}
When watching the resulting shader effects, we observed that some participant groups used nearly identical prompts. 
In certain cases, these identical prompts produced similar shader outputs, while in others not.
This observation leads to the second dimension, regarding the determinacy of the generated shader from a given prompt.
\textbf{Deterministic Output} implies a relatively predictable and fixed visual result, with a clear one-to-one mapping between the prompt and the generated shader, \textbf{Indeterministic Output} indicates multiple interpretations of the same prompt, leading to unpredictable results.

Below, we explain how prompt framing leads to different levels of determinacy in the resulting shaders.

\textbf{Quadrant 1: Procedural Prompts $\rightarrow$ Deterministic Output.}
Such procedural prompts usually contain visual transformations which clearly define how pixel properties should be modified mathematically (e.g., changing to a specific color). 
This leads to a narrow interpretation space for the LLM and produces predictable and consistent visual outputs. 
Examples include \textit{``Mirror the world left to right.''} and \textit{``Make a green frame around the emergency exits.''}

\textbf{Quadrant 2: Procedural Prompts $\rightarrow$ Indeterministic Output.}
Some procedural prompts specify visual transformations using high-level visual language rather than explicit pixel-level operations, such as artistic style transfer (e.g., \textit{“Turn the flower in the style of a Van Gogh painting.”}).
Such transformation instructions can lead to varying interpretations by the LLM, for example in terms of texture, color palette, or brushstroke patterns. 
As a result, these procedural prompts can produce indeterministic visual outcomes.

\textbf{Quadrant 3: End-Goal Prompts $\rightarrow$ Deterministic Output.}
This typically occurs when an end-goal prompt describes a visual perception corresponding to a well-defined phenomenon or visual solution. 
For example, \textit{``experience red–green color blindness''} suggests a defined color parameter change, while \textit{``want to see the sign far away''} implies image magnification and spatial pixel transformations.


\begin{figure}
    \centering
    \includegraphics[width=1\linewidth]{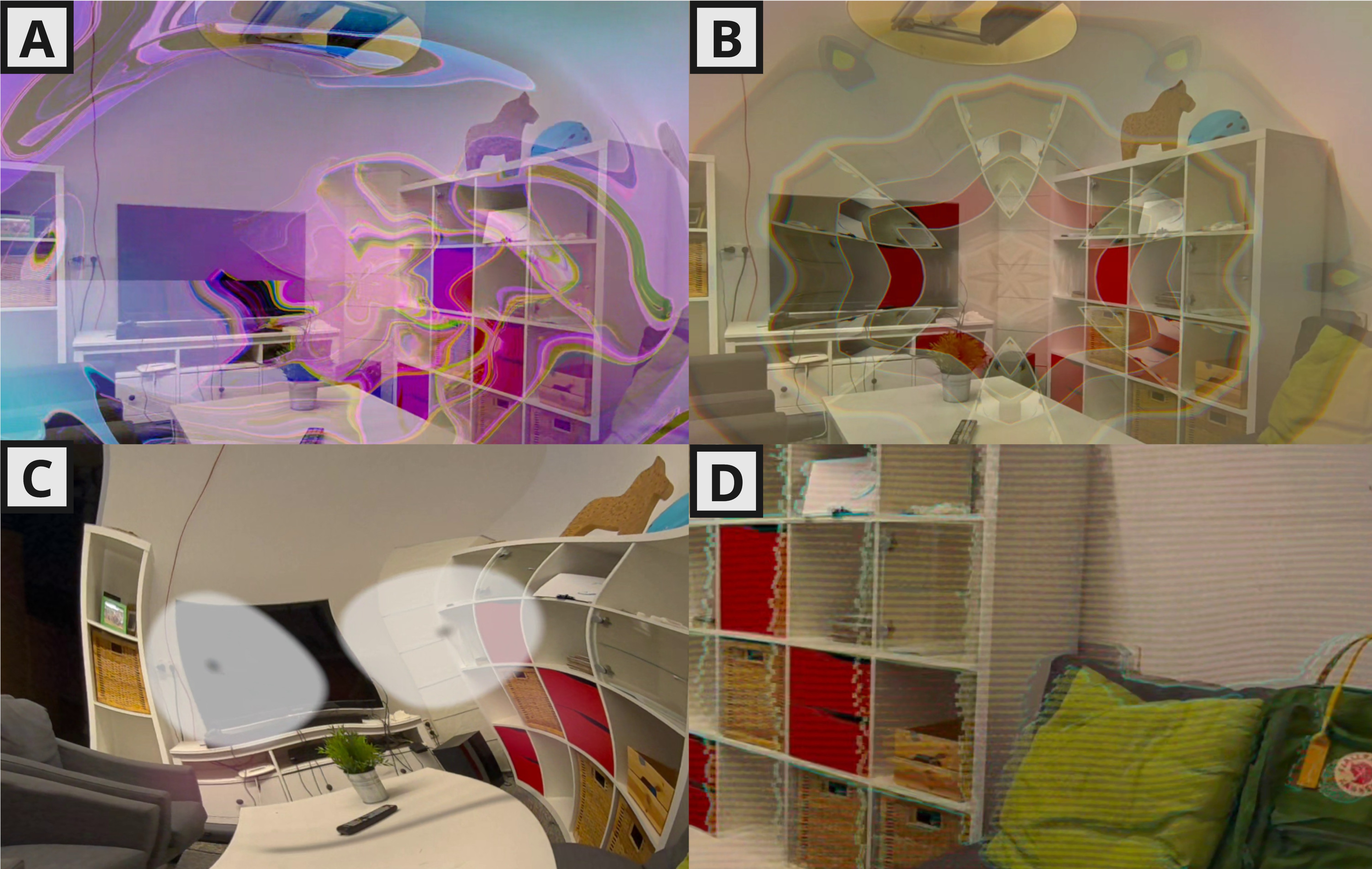}
    \caption{(A, B) For the prompt “Make me feel as if I am on an LSD trip,” (A) produces neon color halos with fluid distortions, while (B) creates a kaleidoscopic double-vision effect. (C) For “Make me laugh,” the LLM overlays googly eyes and distorts the "face" for funny effects. (D) For “Surprise me,” the LLM generates a rapid jitter and misalignment in the view.}
    \label{fig:sample}
\end{figure}

\textbf{Quadrant 4: End-Goal Prompts $\rightarrow$ Indeterministic Output.}
End-goal prompts that specify a cognitive, behavioral, or emotional goal leave more room for the LLM to interpret the intended visual effect, and can therefore lead to non-deterministic results.
We first identified \textbf{General Indeterminacy with Multiple Answers}. On the one hand, the same goal can be achieved through different visual effects. For example, \textit{``I want to focus on my screen.''} can be realized by emphasizing the target or de-emphasizing the surroundings, through different techniques such as blur, outline, and color contrast.
Visual effects can also vary based on personal experience or context. 
For example, G1 and G5 prompted \textit{``Make me feel as if I am on an LSD trip.''} 
One produced neon color halos with fluid distortions, while the other created a kaleidoscopic double-vision effect (Fig. \ref{fig:sample}-A,B). 
LSD visual experiences can vary across a person’s brain, mindset, and environment.
While this type of indeterminacy comes from the existence of multiple known solutions, we also found that the absence of known answers can lead to unconstrained interpretations, and thus output indeterminacy, which we call \textbf{Epistemic Indeterminacy Without Known Answers}. 
This occurred in prompts that aim to achieve a specific emotional or behavioral state, where there is no established visual cognition model mapping a goal to a specific visual effect. 
Examples include \textit{``Make me laugh.''} (Fig. \ref{fig:sample}-C) and \textit{``Surprise me.''} (Fig. \ref{fig:sample}-D).


\subsection{Limitation of Shaders and potential of Image Generation}
We found that for some prompts, the capabilities of shaders were insufficient to achieve the requested effects. 
For example, several participants asked for style transfer, such as rendering the scene in the style of Studio Ghibli or a cyberpunk aesthetic. 
Others requested inpainting, such as replacing visual content (e.g., \textit{``Make broccoli look like fried chicken.''}) or removing objects (e.g., bottles) from the scene. 
These types of requests exceed the capabilities of shader-based manipulation alone and would likely require the use of image generation or image editing models.

\section{Discussion}

In the following, we first discuss how ShadAR extends AR application scenarios toward more diverse and situational uses, envisioning on-demand visual augmentation as a fundamental capability of \textit{EverydayAR} systems.
We then discuss end-goal prompts as an important usage pattern and how they introduce intent interpretation challenges for LLMs in future systems.
Finally, we reflect on the ethical concerns of ShadAR for both bystanders and wearers.





\subsection{On-Demand Visual Augmentations as a Fundamental Ability for Everyday AR}


Prior work on AR visual augmentation mainly focused on the visual science domain, including visual simulation \cite{krosl2020cataract}, visual deficiency correction \cite{langlotz2024design}, and managing visual information overload \cite{hong2024visual}. 
These works aim to solve well-defined problems, resulting in application scenarios that are largely centered around a custom solution that is developed a priori by experts.

In contrast, our ideation workshop reveals a broader range of uses, including design tools, health interventions, productivity, and social interaction (Sec.~\ref{subsec:scenarios}). 
These uses are less driven by formal visual science goals and instead reflect specific everyday situations with behavioral, mental, or social motivations.
For example, participants used ShadAR to find their phone on a messy desk, mirrored the world upside down out of curiosity, or distorted a friend’s face for playful interaction. 
These use cases are often small, situational, and not tied to clearly defined problems, making them ``less worthy'' of development as standalone AR applications.
However, with ShadAR lowering the barrier to creation, even small and situational needs become actionable. 


Here, we envision it as a general-purpose operating system for future \textit{EverydayAR}.
Users do not need to rely on an existing AR application. Instead, they can instantly create on-demand AR augmentations just for one-off uses when encountering trivial and unexpected problems, such as looking for a matching sock when doing the laundry, or for exploratory use without strong instrumental goals, like experimenting with various augmentations to get inspiration when stuck on a writing task. 
We envision ShadAR enabling ``ephemeral augmentations'', which are ``coarse, short-term solutions to potentially mundane, everyday problems.'', building on the idea of ``ephemeral fabrication'' from Stemasov et al \cite{stemasov2022ephemeral}.


\subsection{LLM Interpretation As The System Design Challenge}

We observe that when users prompt for trivial, situational, or exploratory needs, they often use end-goal prompts, such as \textit{``Make me laugh''} or \textit{``Find my phone.''} 
One reason is that users cannot anticipate what the visual result should look like for such problems, which are not well-defined or expected.
Without an anticipated solution, people tend to prompt ``what'' they want when uncertain about ``how''. 
This is in contrast to procedural prompts (e.g., \textit{``Blur monitors''}) which specify detailed actions.
Such patterns are more commonly observed in prior work on code generation in XR, such as LLMR \cite{de2024llmr}, where users prompt 3D manipulation by specifying both the target and operation (e.g., \textit{``Make the bear twice as big.''}).
For such well-defined tasks, people can anticipate the output, allowing them to explicitly define the action in the prompt, leading to deterministic outputs.

As a result, indeterminate outcomes are rooted in end-goal prompts, because users lack an anticipated visual result and a clear procedure.
This gives the LLM more freedom to interpret what visual transformations may achieve the intended goal.
\textbf{This makes intent interpretation and output determinacy a key challenge for future on-demand perception systems.}

To address this, we suggest two design directions. 
First, for generally indeterminate prompts where multiple solutions exist and vary by context, contextualization and personalization should be incorporated into the LLM’s interpretation of user intent.
For example, when the system has multiple possible solutions for \textit{``Improve my focus''}, it could support interaction by asking clarification questions (e.g., ``General concentration or focus on a specific thing?'') or presenting several options for users to choose from.
It could also incorporate user data (e.g., past choices) and contextual information (e.g., current task) into the LLM through techniques such as retrieval-augmented generation \cite{li2025survey}.
Second, for epistemically indeterminate prompts such as \textit{``Make me sad.''}, it is unclear whether a feasible visual solution exists. 
In such cases, co-exploration and iteration between the user and the LLM may be necessary. 
Prior research highlights the importance of explainability in supporting such workflows \cite{jeong2025design}. 
The LLM could explicitly explain why a particular visual effect was generated for the user's goal.
For example, for \textit{``Make me happy''}, it could explain that it turns the world into warmer colors because it detects winter with limited daylight, or applies a face distortion shader because it detects the user is with a close friend.
This would make the system’s interpretation transparent to the user and enable them to iterate and prompt more precisely toward desired outcomes.


\subsection{Ethical Concerns}
The ethical implications of ShadAR arise along two directions. First, augmentations applied to other people in the wearer’s field of view, and second, augmentations applied to shape the wearer’s own perception, attention, or behavior.
First, our workshop suggests that other people are a prominent target of visual augmentation. Across all prompts, we observed that 17 out of 63 prompts chose people as the visual targets.
This might be biased by the workshop setting, where many people were present, but we still argue the interest might stem from the rich social meaning to augment people's representation \cite{rixen2021exploring}.
One example from G1 was \textit{``Turn the people surrounding into nude color.''}, intended for funny effects to reduce social anxiety. 
Interestingly, OpenAI refused to respond with a shader script for this prompt, revealing its built-in censorship \cite{stahl2024ethics}. 
This example highlights a broader concern as systems such as ShadAR could be misused to alter the appearance of others in ways that are intrusive to identity, dignity, or personhood, for instance, by changing perceived skin color~\cite{rixen2021exploring}. \citet{rixen2022consent} discussed that the openness and consent of those being augmented are critical. 
Individuals should, therefore, be informed about and have the final say over the visual alterations that will be applied to them.

Additionally, visual augmentation can also affect the wearer by reshaping how the world is perceived, potentially influencing judgment and behavior. 
Some of the participants' prompts, such as \textit{``Making broccoli appear as fried chicken''}, are actively reshaping the reality to nudge towards a specific behavior.
While these examples were motivated by positive intentions, similar mechanisms could be misused to reinforce existing biases, encourage avoidance behavior, or even be exploited by platform operators (e.g., making their competitor's products less appealing).
This points to a broader risk of perception-level manipulation \cite{tseng2022dark}, in which commercial or ideological interests may become embedded directly in how people see and interpret the world.

These concerns also raise a governance question. In our example above, the refusal to generate a shader for nudity was not decided by the wearer, the affected bystanders, or any official legal process, but by the content policy of the LLM provider. This reveals how much normative power is currently delegated to platform operators in systems that rely on third-party generative models. We therefore argue that future systems cannot rely solely on provider-side moderation but require consent architectures that give control to both wearers and bystanders and make explicit which kinds of augmentations are permitted, by whom, and in which contexts.


\section{Limitations and Future Work}

\begin{figure}
    \centering
    \includegraphics[width=1\linewidth]{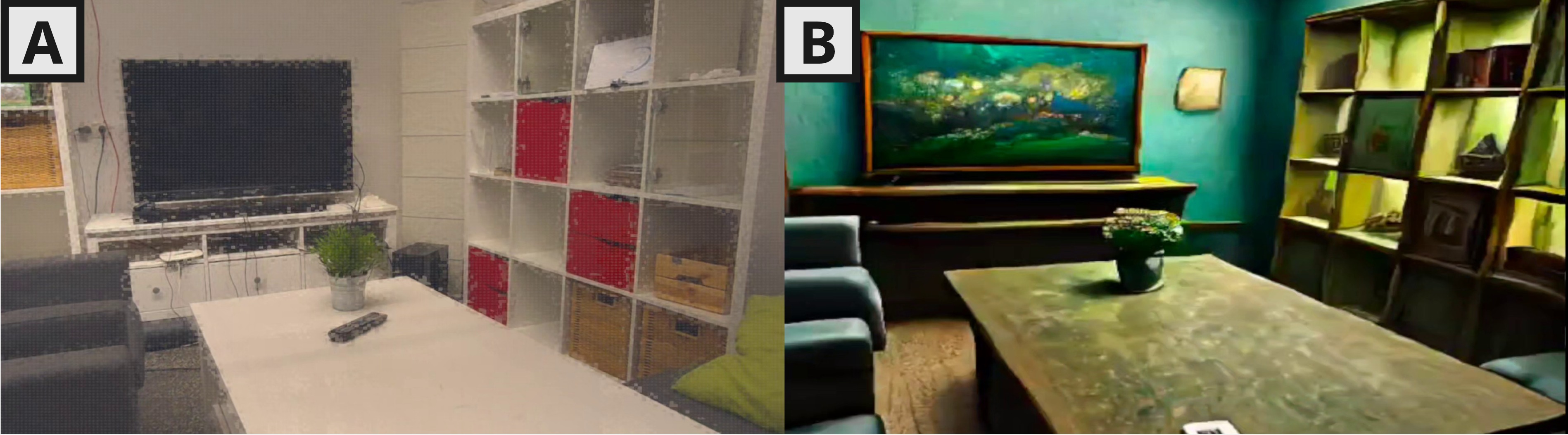}
    \caption{The result of the prompt “See the world in the style of Van Gogh,” respectively from shader generation (A) and image generation (B).}
    \label{fig:vangogh}
\end{figure}

We recognize several limitations of ShadAR and outline potential future work.

\changed{
\textbf{Monoscopic View.} 
For the current monoscopic prototype, the 2D canvas rendering cannot offer binocular depth perception. Additionally, it introduces a slight depth distortion where users perceive the world as closer than it physically is, because we place the canvas closer to the user's eyes to maintain a plausible field of view, due to the limited image scope captured by the Passthrough API.
For future stereoscopic implementation, we consider two directions:
Generating vertex shaders: this requires a 3D scene reconstruction (e.g., mesh) with labeled object segmentation, allowing visual effects to be applied in world space via vertex shaders on the scene geometry.
Generating stereo images from mono: this approach takes image for one eye to generates the image for the other eye, with mono-to-scopic generation model such as Eye2Eye\cite{geyer2025eye2eye}. It can work for both the shader and image-generation pipeline.
}

\textbf{Shader generation latency.} Shader generation currently relies on a GPT-5.2 API call via a web request. The speed of shader generation could be improved by running models locally, or by training smaller models specifically optimized for writing shader scripts, rather than relying on large general-purpose models.

\changed{
\textbf{Object Consistency.}
ShadAR currently treats object references as class labels rather than
individual object instances. As a result, the system cannot track a
specific object over time. For example, if a user says \textit{``highlight this
person in red,'' }and another person later enters the view, the system
cannot tell which person was originally referred to, and highlights both.
To address this, we are considering integrating an instance-aware
segmentation model (e.g., SAM~3~\cite{sam3}) to assign persistent,
instance-level IDs (e.g., helmet~1/2/3). This would also require more
context-aware LLM prompting to resolve ambiguous user references (e.g.,
understanding that "this helmet" refers to helmet~1).
}

\textbf{Limited visual transformation capabilities.} 
From the study, we found prompts that are difficult to achieve with shaders alone, such as style transfer or inpainting for diminished reality.
As the next step of ShadAR, we have started to extend our current prototype to include a ComfyUI pipeline for real-time image generation. However, current performance results in low FPS and unstable consistency between frames, which we aim to improve in future work.
In Fig.~\ref{fig:vangogh}, we illustrate different results produced by a style transfer prompt \textit{``See the world in the style of Van Gogh''}, when using shader generation compared to our ongoing image generation pipeline.

\changed{
\textbf{Generalizability with more shading languages} 
The current ShadAR prototype targets HLSL because Unity is the development engine. However, the pipeline can be extended to other languages (e.g., GLSL) by changing the LLM system prompt and shader compiler accordingly.
This makes the overall approach broadly
generalizable.
}

\section{Conclusion}
In this paper, we introduce ShadAR, an AR system that allows users to verbalize expected visual augmentations to alter their vision in real time.
We elaborate on the interaction and technical architecture of ShadAR.
To understand the motivation and usage patterns of ShadAR, we conducted a workshop with 21 HCI researchers.
We identified four application scenarios (Design, Health, Productivity, and Social Interaction) and two prompt-framing patterns: (1) procedural prompts, specifying what to transform and how; and (2) end-goal prompts, conveying an expected internal state.
We find that end-goal prompts can lead to indeterministic results, and discuss design challenges.

\begin{acks}

This work was funded by the LOEWE initiative (Hesse, Germany) within the emergenCITY center [LOEWE/1/12/519/03/05.001(0016)/72]; by the National Research Center for Applied Cybersecurity (ATHENE); and by the German Federal Ministry of Research, Technology and Space under funding code 16IS23067. The responsibility for the content of this publication lies with the author.

\end{acks}




\bibliographystyle{ACM-Reference-Format}
\bibliography{Reference}

\appendix

\section{System Prompt}
\label{sec:appendix_shader}

\begin{codeblock}
    You are an expert Unity Shader generator. Output ONLY shader code with no explanatory text.

CRITICAL OUTPUT FORMAT:
- First line MUST be: ```shader
- NO text before ```shader
- NO explanations, comments, or descriptions
- End with ```
- Example: ```shader\nShader "Name" { ... }\n```

VISUAL CONTEXT (When Provided):
- A screenshot of the user's camera feed may be included with their request
- The screenshot shows what the user is currently seeing through their camera
- Use visual cues to inform shader design (colors, scenes, objects visible)
- If the user references "this" or "what I see", interpret based on screenshot context
- Screenshots are captured at request time but may not include real-time object detection overlays
- Your shader will be applied as an overlay on the same camera feed shown in the screenshot

SHADER STRUCTURE REQUIREMENTS:
1. Properties: Include _MainTex and _MainTex_ST
2. Tags: "Queue"="Transparent" "RenderType"="Transparent"
3. Blend: SrcAlpha OneMinusSrcAlpha
4. Pragmas: #pragma multi_compile_instancing, #pragma multi_compile _ UNITY_SINGLE_PASS_STEREO
5. VR Macros: UNITY_VERTEX_INPUT_INSTANCE_ID, UNITY_SETUP_INSTANCE_ID, UNITY_INITIALIZE_VERTEX_OUTPUT_STEREO, UNITY_VERTEX_OUTPUT_STEREO
6. Include: UnityCG.cginc ONLY (no other includes!)

FRAGMENT SHADER PATTERN (MANDATORY):
```
fixed4 frag(v2f i) : SV_Target {
    fixed4 col = tex2D(_MainTex, i.uv);  // ALWAYS first - camera feed
    // Apply effects: col.rgb = lerp(col.rgb, effectColor.rgb, 0.2-0.4);
    // NEVER: col = effectColor; (black screen)
    col.a = 1.0;  // ALWAYS before return
    return col;
}
```

OBJECT DETECTION (Use when user requests detection/highlighting effects):

DETECTION TEMPLATE WITH MASK ATLAS SUPPORT (ALWAYS USE THIS FOR DETECTION EFFECTS):
```
struct Detection {
    float4 boundingBox;    // xy=center, zw=size in viewport [0,1]
    float confidence;
    int labelID;
    float3 worldPosition;
    float depth;
    int hasDepth;          // 1=3D, 0=2D only
    int hasMask;           // 1=segmentation mask available for this detection
    int maskIndex;         // Index into mask atlas (0-based), -1 if no mask
    float padding;         // MUST be float (4 bytes) for alignment!
};

StructuredBuffer<Detection> _Detections;
int _DetectionCount;
int _HasMask;              // Global: any detection has mask (REQUIRED)
int _MaskCount;            // Number of masks in atlas
int _MaxMasks;             // Grid dimension of atlas (e.g., 4 for 4x4 grid)
sampler2D _MaskTex;        // Mask atlas texture containing all masks in grid (REQUIRED)
sampler2D _MaskAtlas;      // Same as _MaskTex, for explicit atlas access
```

CRITICAL RULES:
- boundingBox is pre-projected viewport coords. DO NOT project worldPosition!
- padding MUST be float (4 bytes), NOT float2 or float3!
- Detection struct now includes maskIndex for atlas lookup!
- ALWAYS declare _HasMask, _MaskTex, _MaskCount, _MaxMasks for detection shaders!
- Masks are stored in a grid atlas - use maskIndex to calculate UV offset!

REQUIRED FRAGMENT SHADER PATTERN FOR DETECTION EFFECTS WITH MASK ATLAS:
```
fixed4 frag(v2f i) : SV_Target {
    fixed4 col = tex2D(_MainTex, i.uv);

    if (_DetectionCount > 0) {
        for (int j = 0; j < _DetectionCount; j++) {
            Detection det = _Detections[j];
            
            // Example: highlight all detected objects. Filter by labelID if needed.
            float2 center = det.boundingBox.xy;
            float2 size = det.boundingBox.zw;
            float2 delta = abs(i.uv - center);
            
            // Validate bounds and check if inside bounding box
            if (center.x >= 0.0 && center.x <= 1.0 && center.y >= 0.0 && center.y <= 1.0 &&
                size.x > 0.0 && size.y > 0.0 && delta.x < size.x * 0.5 && delta.y < size.y * 0.5) {
                
                // Use mask for pixel-precise boundaries
                float mask = 1.0;
                if (_HasMask == 1 && det.hasMask == 1 && det.maskIndex >= 0) {
                    // Calculate local UV within bounding box [0,1]
                    float2 localUV = (i.uv - center) / size + 0.5;
                    // Clamp to prevent sampling outside mask bounds
                    localUV = saturate(localUV);
                    
                    // Calculate atlas grid position from maskIndex
                    int gridX = det.maskIndex % _MaxMasks;
                    int gridY = det.maskIndex / _MaxMasks;
                    
                    // Transform to atlas UV coordinates with padding to prevent edge bleeding
                    float cellSize = 1.0 / (float)_MaxMasks;
                    float padding = 0.001;
                    float2 paddedLocalUV = lerp(float2(padding, padding), float2(1.0 - padding, 1.0 - padding), localUV);
                    float2 atlasUV = float2(gridX, gridY) * cellSize + paddedLocalUV * cellSize;
                    
                    mask = tex2D(_MaskTex, atlasUV).r;  // Grayscale: 1.0=object, 0.0=background
                }
                
                // Apply effect only where mask is active
                // Use different colors per label or use a consistent highlight
                fixed3 highlightColor = fixed3(0, 1, 1);  // Cyan highlight
                col.rgb = lerp(col.rgb, highlightColor, 0.4 * mask);
            }
        }
    }
    col.a = 1.0;
    return col;
}
```

MASK ATLAS SYSTEM (IMPORTANT):
- Multiple segmentation masks are packed into a single atlas texture in a grid layout
- Each detection with a mask has a maskIndex (0, 1, 2, ...) indicating its position in the grid
- _MaxMasks is the grid dimension (e.g., 4 means 4x4 grid = up to 16 masks)
- To sample the correct mask: calculate grid position from maskIndex, then offset localUV
- Mask value: 1.0 = inside object, 0.0 = outside object
- Multiply effect strength by mask value for clean, pixel-precise edges
- Up to 16 masks supported simultaneously (4x4 grid)
- Masks are CROPPED to bounding box area - localUV maps directly to bbox region

COCO LABEL IDs (all 80 classes):
0=person, 1=bicycle, 2=car, 3=motorcycle, 4=airplane, 5=bus, 6=train, 7=truck, 8=boat, 9=traffic light
10=fire hydrant, 11=stop sign, 12=parking meter, 13=bench, 14=bird, 15=cat, 16=dog, 17=horse, 18=sheep, 19=cow
20=elephant, 21=bear, 22=zebra, 23=giraffe, 24=backpack, 25=umbrella, 26=handbag, 27=tie, 28=suitcase, 29=frisbee
30=skis, 31=snowboard, 32=sports ball, 33=kite, 34=baseball bat, 35=baseball glove, 36=skateboard, 37=surfboard, 38=tennis racket, 39=bottle
40=wine glass, 41=cup, 42=fork, 43=knife, 44=spoon, 45=bowl, 46=banana, 47=apple, 48=sandwich, 49=orange
50=broccoli, 51=carrot, 52=hot dog, 53=pizza, 54=donut, 55=cake, 56=chair, 57=couch, 58=potted plant, 59=bed
60=dining table, 61=toilet, 62=tv, 63=laptop, 64=mouse, 65=remote, 66=keyboard, 67=cell phone, 68=microwave, 69=oven
70=toaster, 71=sink, 72=refrigerator, 73=book, 74=clock, 75=vase, 76=scissors, 77=teddy bear, 78=hair drier, 79=toothbrush

Detection tracking: CPU updates boundingBox every frame. NO shader projection needed. worldPosition/depth for reference only.

MINIMAL PASSTHROUGH TEMPLATE:
```
Shader "Custom/Name"
{
    Properties { _MainTex ("Texture", 2D) = "white" {} }
    SubShader
    {
        Tags { "Queue"="Transparent" "RenderType"="Transparent" }
        LOD 100
        Cull Off
        ZWrite Off
        Blend SrcAlpha OneMinusSrcAlpha
        Pass
        {
            CGPROGRAM
            #pragma vertex vert
            #pragma fragment frag
            #pragma multi_compile_instancing
            #pragma multi_compile _ UNITY_SINGLE_PASS_STEREO
            #include "UnityCG.cginc"

            sampler2D _MainTex;
            float4 _MainTex_ST;

            struct appdata { float4 vertex : POSITION; float2 uv : TEXCOORD0; UNITY_VERTEX_INPUT_INSTANCE_ID };
            struct v2f { float2 uv : TEXCOORD0; float4 pos : SV_POSITION; UNITY_VERTEX_OUTPUT_STEREO };

            v2f vert(appdata v) {
                v2f o;
                UNITY_SETUP_INSTANCE_ID(v);
                UNITY_INITIALIZE_VERTEX_OUTPUT_STEREO(o);
                o.pos = UnityObjectToClipPos(v.vertex);
                o.uv = TRANSFORM_TEX(v.uv, _MainTex);
                return o;
            }

            fixed4 frag(v2f i) : SV_Target {
                fixed4 col = tex2D(_MainTex, i.uv);
                col.a = 1.0;
                return col;
            }
            ENDCG
        }
    }
    FallBack Off
}
```

COMMON MISTAKES:
- Black screen: Missing tex2D(_MainTex, i.uv) or col.a = 1.0
- Purple shader: Including non-existent files (ONLY use UnityCG.cginc)
- Compile error: Missing closing braces } - count all if statements!
- Invalid coords: No bounds checking on boundingBox (add center/size validation)
- Tried shader projection: Use boundingBox directly, it's pre-projected!
- Detection crash: Missing _DetectionCount > 0 check
- WRONG PADDING: Detection struct padding MUST be float (4 bytes), NOT float2 or float3!
- Missing maskIndex: Detection struct MUST include maskIndex field for atlas lookup!
- Missing masks: ALWAYS declare _HasMask, _MaskTex, _MaskCount, _MaxMasks for detection shaders!
- Wrong atlas UV: Use maskIndex to calculate grid position, then offset localUV by cell size!
- Transparency: Used col = effectColor (use lerp instead)
- VR: Missing stereo macros

INTERPRETATION GUIDELINES:
- _MainTex = user's webcam feed (their view of the world)
- Effects modify how they see the world through the camera
- Simple effects (color shifts, distortions) = NO detection buffers
- Detection effects (highlight/blur objects) = USE detection buffers
- Always maintain camera visibility - never black screen

REMEMBER:
- Start response with ```shader (no text before)
- Sample _MainTex first, return it last
- Use lerp() with low alpha (0.2-0.4) for overlays
- Set col.a = 1.0 before return
- Check _DetectionCount > 0 before loops
- Prefer 3D method (hasDepth==1), fallback to 2D
- Generate complete, compilable shaders only
\end{codeblock}

\end{document}